\documentclass[11pt]{article}%
\usepackage{amsfonts}
\usepackage{amsmath}
\usepackage{amssymb}
\usepackage{graphicx}%
\providecommand{\U}[1]{\protect\rule{.1in}{.1in}}
\newtheorem{theorem}{Theorem}
\newtheorem{acknowledgement}[theorem]{Acknowledgement}

\begin{document}

\title{Bohm's Quantum Potential as an Internal Energy}
\author{Glen Dennis\\TPRU, Birkbeck College\\University of London\\London, WC1E 7HX
\and Maurice A. de Gosson\\University of Vienna\\Faculty of Mathematics, NuHAG\\Oskar-Morgenstern-Platz 1, 1090 Vienna
\and Basil J. Hiley\\TPRU, Birkbeck College\\University of London\\London, WC1E 7HX}
\maketitle

\begin{abstract}
We pursue our discussion of Fermi's surface initiated in Dennis, de Gosson and
Hiley and show that Bohm's quantum potential can be viewed as an internal
energy of a quantum system. This gives further insight into the role it played
by the quantum potential in stationary states. It also allows us to provide a
physically motivated derivation of Schr\"{o}dinger's equation for a particle
in an external potential.

\end{abstract}


\section{Introduction}

The time evolution of a quantum system with wavefunction $\psi=\psi
(\mathbf{r},t)$ in physical space, $\mathbb{R}^{3}$, is governed, in
non-relativistic quantum mechanics, by the Schr\"{o}dinger equation
\begin{equation}
i\hbar\frac{\partial\psi}{\partial t}(\mathbf{r},t)=-\frac{\hbar^{2}}%
{2m}\nabla_{\mathbf{r}}^{2}\psi(\mathbf{r},t)+V(\mathbf{r},t)\psi
(\mathbf{r},t). \label{schr1}%
\end{equation}
Writing the wavefunction in polar form $\psi(\mathbf{r},t)=R(\mathbf{r}%
,t)e^{iS(\mathbf{r},t)/\hbar}$, this equation is mathematically equivalent to
the system of real equations
\begin{gather}
\frac{\partial S}{\partial t}(\mathbf{r},t)+\frac{1}{2m}(\nabla_{\mathbf{r}%
}S(\mathbf{r},t))^{2}+Q(\mathbf{r},t)+V(\mathbf{r},t)=0\label{Bohm1}\\
\frac{\partial\rho}{\partial t}(\mathbf{r},t)+\nabla_{\mathbf{r}}\cdot\left(
\rho(\mathbf{r},t)\frac{\nabla_{\mathbf{r}}S}{m}(\mathbf{r},t)\right)  =0.
\label{Bohm2}%
\end{gather}
Equation (\ref{Bohm1}) can be regarded as a Hamilton--Jacobi equation derived,
not from the classical Hamiltonian, but from the $\psi$-dependent Hamiltonian%
\[
H^{\psi}(\mathbf{r},\mathbf{p},t)=\frac{1}{2m}|\mathbf{p}|^{2}+Q(\mathbf{r}%
,t)+V(\mathbf{r},t)\text{.}%
\]
Equation (\ref{Bohm2}) can be viewed as a continuity equation for the
probability $\rho(\mathbf{r,t)}=R\mathbf{^{2}(r,t)}$. We are going to show
that the additional term $Q(\mathbf{r},t)$ (the \textquotedblleft quantum
potential\textquotedblright) can be interpreted as an internal energy
associated with a certain region of phase space, absent in classical
mechanics, but arising in quantum mechanics from the uncertainty principle. In
order to explain how this internal energy arises we must first return to
consider arguments outlined in our recent paper \cite{degohi14} where we
investigated the consequences of Fermi's idea \cite{Fermi} which associated
every quantum state with a certain geometric curve or, more generally, a
hypersurface in phase space.

\section{The Fermi Hamiltonian}

Consider a wavefunction $\psi_{0}(\mathbf{r})=R_{0}(\mathbf{r})e^{iS_{0}%
(\mathbf{r})/\hbar}$, which we assume represents a particle with mass $m$ in
physical space $\mathbb{R}^{3}$ at the initial time $t=0$; here $\mathbf{r}%
=(x,y,z)$ is the position vector. At this point we do not consider an explicit
time dependence of $\psi_{0}$. We assume that $R_{0}(\mathbf{r})>0$ and that
$R_{0}$ is twice continuously differentiable, and that the phase $S_{0}$ is
real and continuously differentiable. It is easily verified that the function
$\psi_{0}$ satisfies the second-order partial differential equation%
\begin{equation}
\left[  \frac{1}{2m}\left(  -i\hbar\nabla_{\mathbf{r}}-\nabla_{\mathbf{r}%
}S_{0}(\mathbf{r})\right)  ^{2}+\frac{\hbar^{2}}{2m}\frac{\nabla_{\mathbf{r}%
}^{2}R_{0}(\mathbf{r})}{R_{0}(\mathbf{r})}\right]  \psi_{0}(\mathbf{r})=0;
\label{fermi5}%
\end{equation}
this can be done by a direct calculation, or by noting that a change of gauge
making $S_{0}=0$, immediately gives%

\[
\frac{-\hbar^{2}}{2m}\left[  \nabla_{\mathbf{r}}^{2}-\frac{\nabla_{\mathbf{r}%
}^{2}R_{0}(\mathbf{r})}{R_{0}(\mathbf{r})}\right]  R_{0}(\mathbf{r})=0.
\]
We can rewrite equation (\ref{fermi5}) more concisely as
$\widehat{H_{\mathrm{F}}}\psi=0$ where $\widehat{H_{\mathrm{F}}}$ is the
\textquotedblleft Fermi operator\textquotedblright\
\begin{equation}
\widehat{H_{\mathrm{F}}}=\frac{1}{2m}\left(  -i\hbar\nabla_{\mathbf{r}}%
-\nabla_{\mathbf{r}}S_{0}(\mathbf{r})\right)  ^{2}-Q_{0}(\mathbf{r}).
\label{hf}%
\end{equation}
The function $Q_{0}$ is given by%
\begin{equation}
Q_{0}(\mathbf{r})=-\frac{\hbar^{2}}{2m}\frac{\nabla_{\mathbf{r}}^{2}%
R_{0}(\mathbf{r})}{R_{0}(\mathbf{r})}. \label{qpfermi1}%
\end{equation}
One immediately recognizes that $Q_{0}$ is the quantum potential at time
$t=0$. The operator $\widehat{H_{\mathrm{F}}}$ is the quantization (in any
reasonable quantization scheme) of the Hamiltonian function%
\begin{equation}
H_{\mathrm{F}}(\mathbf{r},\mathbf{p})=\frac{1}{2m}|\mathbf{p}-\nabla
_{\mathbf{r}}S_{0}(\mathbf{r})|^{2}-Q_{0}(\mathbf{r}). \label{hfcl}%
\end{equation}
Let us consider the energy hypersurface%
\begin{equation}
\Sigma_{\mathrm{F}}:H_{\mathrm{F}}(\mathbf{r},\mathbf{p})=0, \label{1}%
\end{equation}
and assume that this hypersurface is the boundary of a phase space set
$\Omega_{\mathrm{F}}$. Following Fermi, we can then identify $\Omega
_{\mathrm{F}}$ with the quantum particle described by the wavefunction
$\psi_{0}$. One can show that this identification is compatible with the
uncertainty principle in the following sense: in quantum mechanics, the notion
of a particle existing at a point in phase space does not make sense.

The set $\Omega_{\mathrm{F}}$ may therefore be viewed as the \textquotedblleft
blow-up\textquotedblright\ of such a point, in fact the smallest entity
unfolded from a point allowed by the uncertainty principle. This blow-up
requires energy, and this energy is the quantum potential $Q_{0}$. We view it
as an \textit{internal energy} associated with the quantum particle, whose
total energy is thus given by
\begin{equation}
E=E_{\mathrm{kin}}+Q_{0}+E_{\mathrm{pot}}. \label{energy}%
\end{equation}
Note that both $E_{\mathrm{kin}}$ and $Q_{0}$ are \emph{internal energies}, as
opposed to $E_{\mathrm{pot}}$ which is energy coming from an external source.

The case of a real bound quantum state is particularly instructive and will be
illustrated in the next section in the case of the harmonic oscillator. Assume
$S_{0}=0$ and that\
\[
\left[  -\frac{\hbar^{2}}{2m}\nabla_{\mathbf{r}}^{2}+V(\mathbf{r})\right]
\psi_{0}(\mathbf{r})=E\psi_{0}(\mathbf{r}).
\]
Using (\ref{hf}) we also have%
\[
\left[  -\frac{\hbar^{2}}{2m}\nabla_{\mathbf{r}}^{2}-Q_{0}(\mathbf{r})\right]
\psi_{0}(\mathbf{r})=0.
\]
Hence, by subtracting these equations, we find the total energy is given by%
\begin{equation}
E=V(\mathbf{r})+Q_{0}(\mathbf{r}). \label{evq}%
\end{equation}
It follows that the classical force $F_{\mathrm{c}}=-\nabla_{\mathbf{r}%
}V(\mathbf{r})$ and $\mathbf{F}_{\mathrm{Q}}=-\nabla_{\mathbf{r}}%
Q(\mathbf{r})$ sum up to zero: $F_{\mathrm{c}}+F_{\mathrm{Q}}=0$. This is
perfectly in accordance with the fact that in Bohm's theory of quantum motion,
the particle in a bound real state is at rest since $E_{\mathrm{kin}}=0$. Let
us probe this counter-intuitive result further.

\section{Example: the Isotropic Harmonic Oscillator}

Let us illustrate the above conclusions with a simple but instructive example.
Choose for $\psi_{0}$, the coherent state
\[
\psi_{0}(\mathbf{r})=e^{-m\omega|\mathbf{r}|^{2}/2\hbar}%
\]
where $|\mathbf{r}|^{2}=x^{2}+y^{2}+z^{2}$. A straightforward calculation
yields%
\[
Q_{0}(\mathbf{r})=-\frac{1}{2}m\omega^{2}|\mathbf{r}|^{2}+\frac{3\omega\hbar
}{2}.
\]
Hence the Fermi operator is $\widehat{H}_{F}=\widehat{H}_{0}-\frac{3}{2}%
\omega\hbar$ where
\[
\widehat{H}_{0}=-\frac{\hbar^{2}}{2m}\nabla_{\mathbf{r}}^{2}+\frac{1}%
{2}m\omega^{2}|\mathbf{r}|^{2}.
\]
One thus recovers the fact that $\psi_{0}$ is the ground state of the
three-dimensional isotropic harmonic oscillator. The corresponding Fermi
function is
\[
H_{F}=\frac{1}{2m}|\mathbf{p}|^{2}+\frac{1}{2}m\omega^{2}|\mathbf{r}%
|^{2}-\frac{3\omega\hbar}{2}%
\]
with $|\mathbf{p}|^{2}=p_{x}^{2}+p_{y}^{2}+p_{z}^{2}$; the Fermi hypersurface
$\Sigma_{\mathrm{F}}$ is here the constant energy set
\[
H_{0}=\frac{1}{2m}|\mathbf{p}|^{2}+\frac{1}{2}m\omega^{2}|\mathbf{r}%
|^{2}=\frac{3\omega\hbar}{2}%
\]
for the classical Hamiltonian $H_{0}$. The Bohm momentum $\mathbf{p}%
=\nabla_{\mathbf{r}}S$ is zero since $\psi_{0}$ is real, and the state's
internal quantum potential energy is
\begin{equation}
Q(\mathbf{r})=\frac{3\omega\hbar}{2}-\frac{1}{2}m\omega^{2}|\mathbf{r}|^{2}.
\label{qr}%
\end{equation}
This is just the ground state energy $3\omega\hbar/2$ minus the potential
energy. Equivalently, all the energy $3\omega\hbar/2$ is obtained by adding
the classical and quantum potential energy. Observe that the quantum force is
here
\[
\mathbf{F}_{\mathrm{Q}}=-\nabla_{\mathbf{r}}Q(\mathbf{r})=m\omega
^{2}\mathbf{r}%
\]
whereas the classical force is $\mathbf{F}_{\mathrm{C}}=-m\omega^{2}%
\mathbf{r}$ (it is a restoring force, directed towards the equilibrium
position), and we thus have $\mathbf{F}_{\mathrm{Q}}+\mathbf{F}_{\mathrm{C}%
}=0$.

\section{Stationary States in General\label{sec:other}}

The counter-intuitive result of a stationary particle in the ground state of
the harmonic oscillator is not unique. In fact it is clear that a particle in
any stationary state described by a real wave function will have zero kinetic
energy and will therefore not be moving.

Consider an even simpler case of a particle trapped between infinitely large
confinement potentials. A straightforward calculation will show that a
stationary particle will occur for \emph{all} energy eigenstates.
Einstein~\cite{ae53}, himself, used this example as an objection to the whole
approach, claiming that it violated physical intuition. He required the
particle to be moving back and forth within the box. If that is the preferred
intuition, then there is the problem of how the particle goes through the
nodes of the wave functions of the higher energy states. For at the nodes, the
quantum potential becomes infinitely repulsive and therefore conservation of
energy would be violated if the particles were actually
oscillating~\cite{dbbh85}.

What our model is telling us is that, in the quantum domain, we must give up
the idea that a particle is represented as a point in phase space. As we
remarked earlier, the blow up requires energy and this energy comes from the
particle itself -- it comes from its kinetic energy. In the extreme case of a
particle described by a real wave function, all the kinetic energy is
transferred into quantum potential energy, the remaining rest mass is absorbed
into the rest of the atom. This situation is reminiscent of the photon where
the whole quantum of energy is absorbed by the atom, thereby completely losing
its identity. There is a difference, however, in that a lepton cannot lose its
identity owing to lepton number conservation.

In the reverse process, the photon emerges at an atomic transition associated
with the process of emission. However, the electron only emerges with its full
kinetic energy when it escapes the Coulomb potential. This brings out one of
the differences in the behaviour of a photon and a lepton (particle), a
difference that forces us to treat the electromagnetic field in a different
way~\cite{dbbhpk87}. Furthermore it illustrates that the concept of a particle
in Bohm's theory is very different from that of a classical
particle~\cite{ppiibh14}.

\section{Derivation of Schr\"{o}dinger's Equation}

Let us now return to our general discussion and suppose the wavefunction
depends on time under the action of some potential $V$ (possibly itself
time-dependent); the internal energy also becomes time-dependent, and the
total Hamiltonian function is thus%
\begin{equation}
H(\mathbf{r},\mathbf{p},t)=\frac{p^{2}}{2m}+Q(\mathbf{r},t)+V(\mathbf{r},t)
\label{3}%
\end{equation}
where $Q(\mathbf{r},t)$ is the quantum potential at time $t$, derived as
above. If we believe that, as in classical physics, the motion of the particle
represented by the wavefunction is determined by the action, we see that the
latter is a solution of the Hamilton--Jacobi equation%
\begin{equation}
\frac{\partial S}{\partial t}+\frac{1}{2m}(\nabla_{\mathbf{r}}S)^{2}%
+Q(\mathbf{r},t)+V(\mathbf{r},t)=0. \label{4}%
\end{equation}

Let us now assume, with Born, that $\rho=\psi^{\ast}\psi=R^{2}$ represents a
probability density; this is consistent with Gleason's theorem \cite{Gleason}.
One way of interpreting Gleason's theorem is to view it as a derivation of the
Born rule from fundamental assumptions about quantum probabilities, guided by
quantum theory, in order to assign consistent and unique probabilities to all
possible measurement outcomes. Gleason proved that there is no alternative to
the Born rule for Hilbert spaces of dimension of at least three. Introducing
the associated probability current
\begin{equation}
\mathbf{j}=\rho\frac{\nabla_{\mathbf{r}}S}{m}, \label{5}%
\end{equation}
conservation of probability is equivalent to the continuity equation%
\begin{equation}
\frac{\partial\rho}{\partial t}+\nabla_{\mathbf{r}}\cdot\mathbf{j}=0;
\label{6}%
\end{equation}
that is to the continuity equation%
\begin{equation}
\frac{\partial\rho}{\partial t}+\nabla_{\mathbf{r}}\cdot\left(  \rho
\frac{\nabla_{\mathbf{r}}S}{m}\right)  =0. \label{7}%
\end{equation}

Summarizing, the time-evolution of the phase $S$ and the amplitude $R$ of the
wavefunction is determined by the system of coupled partial differential
equations (\ref{4}) and (\ref{7}). As is well-known (see \textit{e.g.} Bohm
and Hiley \cite{bohipilot}, Holland \cite{Holland}), this system is equivalent
to the Schr\"{o}dinger equation%
\[
i\hbar\frac{\partial\psi}{\partial t}(\mathbf{r},t)=-\frac{\hbar^{2}}%
{2m}\nabla_{\mathbf{r}}^{2}\psi(\mathbf{r},t)+V(\mathbf{r},t)\psi
(\mathbf{r},t).
\]
Thus we have shown that if we take the internal energy (quantum potential)
into account, then the evolution of the wavefunction is governed by the
Schr\"{o}dinger equation. What happens if we ignore this internal energy?
Then, as one of us has shown in \cite{ICP}, the wavefunction will satisfy the
\emph{nonlinear} partial differential equation%
\[
i\hbar\frac{\partial\psi}{\partial t}(\mathbf{r},t)=-\left(  \frac{\hbar^{2}%
}{2m}\nabla_{\mathbf{r}}^{2}+Q(\mathbf{r},t)\right)  \psi(\mathbf{r}%
,t)+V(\mathbf{r},t)\psi(\mathbf{r},t).
\]

\section{Relation to the Uncertainty Principle}

In \cite{de03-2} one of us introduced the notion of \textquotedblleft quantum
blob\textquotedblright, which is the image of a phase space ball with radius
$\sqrt{\hbar}$ by a linear symplectic transformation, and their study was
pursued in \cite{Birk,blobs}. A quantum blob is an ersatz for the awkward
notion of quantum cell from thermodynamics, and enjoys the pleasant property
of symplectic symmetry; its introduction was motivated by the
Robertson--Schr\"{o}dinger (RS) inequalities, and a quantum blob can be viewed
as the minimum uncertainty set in phase space that is compatible with these
inequalities. Now, the RS inequality
\[
\Delta x^{2}\Delta p_{x}^{2}+\operatorname*{Cov}(x,p_{x})^{2}\geq\frac{1}%
{4}\hbar^{2}%
\]
(and similar relations for the other variables) are expressed in terms of the
variances $\Delta x^{2},\Delta p^{2}$ and the covariance $\operatorname*{Cov}%
(x,p_{x})$; these are conventional measures of spreading, without any precise
physical meaning. For instance, Hilgevoord and Uffink note in
\cite{hi02,hiuf85bis} that variances only give an adequate physical
measurement of the spread of a wavefunction when the probability density is
nearly Gaussian. In the present paper, we have associated a quantum system
with a much more natural notion, that of Fermi set $\Omega_{\mathrm{F}}$,
which is \emph{canonically} associated with the state $\psi$. We have shown in
\cite{degohi14} that in the case of one degree of freedom, where
$\Omega_{\mathrm{F}}$ becomes a phase space surface, we have
$\operatorname*{area}(\Omega_{\mathrm{F}})\geq\frac{1}{2}h$, which is
compatible with the RS inequalities (de Gosson \cite{de02-2,de03-2,go09}, de
Gosson and Luef \cite{golu10}); for an arbitrary number of degrees of freedom
we conjecture that this condition on areas should be replaced with
$c(\Omega_{\mathrm{F}})\geq\frac{1}{2}h$ for some symplectic capacity $c$.
This property would imply that every Fermi set $\Omega_{\mathrm{F}}$ contains
\textit{de facto} \cite{blobs,golu10} a quantum blob, in accordance with the
uncertainty principle. This allows us to define $\Omega_{\mathrm{F}}$ as the
quantum-mechanical counterpart of a classical point-like particle,
generalising the discussion in Section \ref{sec:other}.

Let us illustrate this on the example of the ground state of the
three-dimensional isotropic harmonic oscillator. Here the Fermi set is the
phase space ellipsoid%
\[
\Omega_{\mathrm{F}}:\frac{1}{2m}|\mathbf{p}|^{2}+\frac{1}{2}m\omega
^{2}|\mathbf{r}|^{2}\leq\frac{3\omega\hbar}{2}.
\]
An easy calculation shows that the intersection of $\Omega_{\mathrm{F}}$ with
any of the conjugate variable planes $x,p_{x}$, $y,p_{y}$, and $z,p_{z}$ is an
ellipse with area $3h/2$. It follows \cite{eggs} that the symplectic capacity
of $\Omega_{\mathrm{F}}$ is
\[
c(\Omega_{\mathrm{F}})=\frac{3h}{2}.
\]
Hence $\Omega_{\mathrm{F}}$ contains a \textquotedblleft quantum
blob\textquotedblright. This statement can actually be refined by using the
notion of Ekeland--Hofer capacity, which allows the classification of phase
space ellipsoids using the action integrals of the periodic orbits on the
boundary of the Fermi set $\Omega_{\mathrm{F}}$; we will develop this approach
in a forthcoming paper.

\begin{acknowledgement}
Maurice de Gosson has been supported by the Austrian Research Agency FWF
(Projektnummer P20442-N13).
\end{acknowledgement}

\end{document}